\documentclass[aps,prd,twocolumn,nofootinbib]{revtex4}
\usepackage{epsfig}
\usepackage{amsmath}
\usepackage{graphicx}
\newcommand{\beq}{\begin{equation}}
\newcommand{\eeq}{\end{equation}}
\newcommand{\bqa}{\begin{eqnarray}}
\newcommand{\eqa}{\end{eqnarray}}
\newcommand{\nnb}{\nonumber\\}

\begin{document}

\title{Understanding the cross section of $e^+e^-\to \eta J/\psi$
process via nonrelativistic QCD}
\author{Cong-Feng Qiao}
\email{qiaocf@ucas.ac.cn}
\affiliation{School of Physics, University of Chinese Academy of Sciences \\
YuQuan Road 19A, Beijing 100049, China} \affiliation{Collaborative
Innovation Center for Particles and Interaction, USTC, HeFei 230026,
China}

\author{Rui-Lin Zhu}
\email{zhuruilin09@mails.ucas.ac.cn}
\affiliation{School of Physics, University of Chinese Academy of Sciences \\
YuQuan Road 19A, Beijing 100049, China} \affiliation{Nuclear Science
Division, Lawrence Berkeley National Lab, Berkeley, CA 94720, USA}


\begin{abstract}

Motivated by the large cross section of $e^+e^-\to \eta J/\psi$
process measured  by the BESIII  and Belle Collaborations recently,
we evaluate this process at $\mathcal{O}(\alpha_s^4)$ accuracy in
the framework of nonrelativistic QCD. We find that the cross section
at the center-of-mass energy  $\sqrt{s}=4.009$ {GeV} is
$34.6\mathrm{pb}$, which is consistent with the BESIII data.
The comparison with the
Belle data in the region from $4.0\ \mathrm{GeV}$ to $5.3\
\mathrm{GeV}$ is also presented. Concerning the $\eta$ and
$\eta^\prime$ mixing and the potential gluonium component of
$\eta^\prime$, we also estimate the rate of $e^+e^-\to \eta^\prime
J/\psi$ process, which can be checked within the updated Belle
data.

\vspace{0.3cm}

\end{abstract}

\maketitle
\section{Introduction}
The quarkonium spectroscopy has represented a unique inspection window on
strong interaction for more than thirty years, and keeps on
attracting great interests from both theorists and experimentalists
up to now. It is well-known that the Constituent Quark Model (CQM)
predicts perfectly the spectroscopy of charmonium in the scope below
$D\bar{D}$ threshold, but still has to rise to the challenge in the region
above the threshold \cite{HQW}. In this region, a plethora of new states
confirmed by different experiments are in or beyond the
spectroscopy anticipated by the CQM.
On one hand, conventional resonances such as $\psi(3770)$, $\psi(4040)$,
$\psi(4160)$, and $\psi(4415)$, whose quantum numbers can be
predicted by CQM, strongly couple to open charm final states and
then have broad widths\cite{kzhu}. On the other hand, many newly
observed charmonium-like states with $J^{PC}=1^{--}$, the Y(4260),
Y(4360), and Y(4660), exhibit an exotic nature, which couple to
hidden charm final states with large partial widths \cite{Belle}. This suggests
that these states may have different structures with respect to their
conventional descriptions within the CQM.

By studing the hadronic transition to a lower charmonium state like
$J/\psi$ and a light meson like  $\eta$, we can investigate the
properties of those conventional or unconventional states in
question. The CLEO collaboration measured the cross section of
$e^+e^-\to \eta J/\psi$ as $15^{+5}_{-4}\pm8 $ {pb} at
$\sqrt{s}=4.12-4.20$ {GeV} eight years ago \cite{Coan:2006rv}.
Recently, the BESIII Collaboration has measured the cross section
of $e^+e^-\to \eta J/\psi$ at $\sqrt{s}=4.009$ {GeV} \cite{BesIII}.
The result is $\sigma(e^+e^- \to \eta J/\psi)(\sqrt{s}=4.009\
\text{GeV})= (32.1\pm2.8\pm1.3)\ \mathrm{pb}$, which is a large
value. And the Belle Collaboration has also measured the cross
section of the same channel by scanning the center-of-mass energy
from $3.8\ \mathrm{GeV}$ to $5.3\ \mathrm{GeV}$ \cite{Belle2}. They
show us that the cross sections are around $70\ \mathrm{pb}$ and
$50\ \mathrm{pb}$ when the center-of-mass energy are near the
$\psi(4040)$ and $\psi(4160)$ peaks respectively. We can imagine
that the resonance effects play an important role in those energy
areas. The pioneering work studying the open charm effects in this
process can be found in Ref.~\cite{Wang:2011yh}. Therein the cross
section of $e^+e^-\to \eta J/\psi$ is calculated via virtual D meson
loops at $\sqrt{s}$ from $3.7\ \mathrm{GeV}$ to $4.3\ \mathrm{GeV}$.
And, in those regions, the results are in agreement with the experimental
data with suitable choices of input parameters.

In this paper, we adopt the nonrelativistic QCD (NRQCD) scheme and
calculate the cross section of $e^+e^-\to \eta J/\psi$ at $\sqrt{s}$
from $4.0\ \mathrm{GeV}$ to $5.3\ \mathrm{GeV}$. It is known that
higher-order corrections to the expansion  in the strong coupling constant $\alpha_s$  can
solve the discrepancy problem in double charmonium production at the
B-factory \cite{exp,theore,NLO,He:2007te}. In this paper the cross section of
$e^+e^-\to \eta^{(\prime)} J/\psi$ is investigated at order of
$\mathcal{O}(\alpha_s^4)$. Our aim is to understand whether the data can be
explained at order  $\mathcal{O}(\alpha_s^4)$.

Hereafter, $\eta$ and $\eta^\prime$ are treated with the Light-Cone
(LC) approach while $J/\psi$ is treated with NRQCD. The $\eta$ and
$\eta^\prime$ mixing effect is also important, as has been widely
discussed in, e.g.
 in Refs.~\cite{Bramon:1997va,decay1,decay2,decay3,prod1}, and will be
taken into account in our considerations. $\eta$ can be described on
either flavor octet-singlet or quark-flavor basis, however for
$\eta^\prime$ the gluonium content may have a role in the interaction.
According to the conservation of parity and charge conjugation, the
leading-order in $\alpha_s$ expansion for $e^+e^-\to \eta^{(\prime)}
J/\psi$ comes from the diagrams where two gluons transit to
$\eta^{(\prime)}$. In this case, the color-singlet Fock state of
$J/\psi$ dominates the contribution.

This paper is organized as follows: after the Introduction, the
Sec.~\ref{sec-form} contains the formulae used in our analysis.
Therein we introduce the LC distribution amplitude for each
component of the $\eta^{(\prime)}$  meson.  The effective
$\eta^{(\prime)} g g$ vertex is employed within the LC approach. Our
numerical results for the cross section are given in
Sec.~\ref{sec-num}, where the comparison with the BESIII and Belle
data are also presented. The last section is reserved to
conclusions.

\section{production mechanism\label{sec-form}}

In the literature, two schemes exist to describe the $\eta$ and
$\eta^\prime$ mixing \cite{Feldman}:
\begin{itemize}
\item with respect to the flavor octet and singlet bases, where the  basis
vectors are $\eta_8=(u\bar{u}+d\bar{d}-2s\bar{s})/\sqrt{6}$ and
$\eta_0=(u\bar{u}+d\bar{d}+s\bar{s})/\sqrt{3}$. Then the states can
be decomposed as
\begin{eqnarray}
\left (
\begin{array}{c}
|\eta \rangle \\
|\eta^{\prime} \rangle \\
\end{array}
\right ) = \left (
\begin{array}{cc}
\cos \theta & -\sin \theta \\
\sin \theta & \cos \theta \\
\end{array}
\right ) \left (
\begin{array}{c}
|\eta_8 \rangle \\
|\eta_0 \rangle \\
\end{array}
\right )  \; ;\label{mixsu3}
\end{eqnarray}

\item with respect to the quark-flavor (QF) bases, where the  basis
vectors are $\eta_q=(u\bar{u}+d\bar{d})/\sqrt{2}$ and
$\eta_s=s\bar{s}$, then
\begin{eqnarray}
\left (
\begin{array}{c}
|\eta \rangle \\
|\eta^{\prime} \rangle \\
\end{array}
\right ) = \left (
\begin{array}{cc}
\cos \phi & -\sin \phi \\
\sin \phi & \cos \phi \\
\end{array}
\right ) \left (
\begin{array}{c}
|\eta_q \rangle \\
|\eta_{s} \rangle \\
\end{array}
\right )  \; . \label{mixqf}
\end{eqnarray}
\end{itemize}

In what follows, we perform our calculation using
the QF bases,  and the latest value of the mixing angle measured by
the KLOE Collaboration is $(41.5\pm0.3\pm0.7\pm0.6)^0$ \cite{KLOE}.
Using the definition in (\ref{mixqf}), we can get the corresponding
decompositions for $\eta$ and $\eta^{\prime}$
\begin{eqnarray}\label{ds}
    &&|\eta\rangle=\cos \phi \;|\eta_q\rangle-\sin \phi \;|\eta_s\rangle\;,\\
    &&|\eta^\prime\rangle=\sin \phi \;|\eta_q\rangle+\cos \phi\;
    |\eta_s\rangle\;.
\end{eqnarray}

In the above decomposition, we still do not consider the gluonium
component, which, however, can be important in
$\eta^{\prime}$\cite{cgge}. Hence we should take  the gluonium
component into account in the analysis for $\eta^{\prime}$. On the other hand,
the QCD sum rules has told us that the gluonium couples
to $\eta$ much more weakly than to $\eta^{\prime}$ \cite{cgge}. In this case we will
adopt the traditional view and ignore the gluonium component in the
$\eta$ meson. Based on this argument, an additional mixing angle
$\phi_G$ is introduced, and the gluonium component is defined as
$|\eta_g\rangle =|gg\rangle$, then the $\eta^\prime$ state is
reparametrized
 as
\begin{eqnarray}\label{ds}
|\eta^\prime\rangle=\cos \phi_G (\sin \phi \;|\eta_q\rangle+\cos
\phi \;|\eta_s\rangle)+\sin \phi_G \;|\eta_g\rangle\;.
\end{eqnarray}

Next we turn to the LC distribution amplitudes of those components.
 The
LC distribution amplitude of the $\eta_q$ or $\eta_s$ component in
$\eta$ can be expanded in Gegenbauer polynomials \cite{Feldman}
\begin{equation}\label{gegen}
   \Phi(x,\mu)=6x\bar x(1+\sum_{n=1}^\infty B_{2n}(\mu)\;
   C_{2n}^{3/2}(x-\bar x))\;,
\end{equation}
where $x$ and $\bar x = 1 - x$ are the momentum fractions of the two
partons inside the $\eta_{q,s}$ component respectively. The scale
dependence of $B_n(\mu)$ at leading-order logarithmic accuracy can
be written as
\begin{equation}\label{gegen}
   B_n(\mu)=\left(\frac{\alpha_s(\mu)}{\alpha_s(\mu_0)}\right)^{
   \frac{\gamma_n}{\beta_0}} B_n(\mu_0)\;,
\end{equation}
where the anomalous dimension reads as
\begin{equation}\label{gegen}
   \gamma_n=4C_F
   (\psi(n+2)+\gamma_E-\frac{3}{4}-\frac{1}{2(n+1)(n+2)})\,,
\end{equation}
being $\psi(n)$  the digamma function. Here $\mu_0 $ is the
typical hadronic energy scale below which non-perturbative evolution
takes place. And all those contributions are summed up
into the factor $B_n(\mu_0)$. In the following calculation,
$B_2(1\mathrm{GeV})=0.2$ is adopted, which is determined from QCD
Sum Rules \cite{Ball}.

For the $\eta^\prime$ state, the mixing effect between quark  and
gluonium components should be taken into account. Then we take the
following form for the LC distribution amplitudes  \cite{etagg}
\begin{widetext}
\begin{eqnarray}
 && \Phi^{(q,s)} (x, \mu) = 6 x \bar x \left \{ 1 +
\left [  B^{(q,s)}_2(\mu_0) \left ( \frac{\alpha_s (\mu^2)}{\alpha_s
(\mu_0^2)} \right )^{\frac{48}{81}} - \frac{B^{(g)}_2(\mu_0)}{90}
\left ( \frac{\alpha_s (\mu^2)}{\alpha_s (\mu_0^2)} \right
)^{\frac{101}{81}} \right ]  C_2^{3/2}(x-\bar x) + \cdots \right \}
\;,\nonumber
 \\
 && \Phi^{(g)} (x, \mu) = x \bar x (x -  \bar x)
\left [ 16 B^{(q,s)}_2(\mu_0) \left ( \frac{\alpha_s
(\mu^2)}{\alpha_s (\mu_0^2)} \right )^{\frac{48}{81}} + 5
B^{(g)}_2(\mu_0) \left ( \frac{\alpha_s (\mu^2)}{\alpha_s (\mu_0^2)}
\right )^{\frac{101}{81}} \right ] + \cdots\;. \label{eq:gg}
\end{eqnarray}
\end{widetext}

According to the NRQCD scheme\cite{Bratten,Qiao:2009}, the physical
state $J/\psi$ is expanded in the Fock space according to powers of  $v$, the
relative velocity of the heavy quark in the quarkonium. The
expansion can be written as
\begin{equation}
|J/\psi\rangle=\mathcal{O}(1)|c\bar{c}[^3S_1^{(1)}]\rangle +
\mathcal{O}(v)|c\bar{c}[^3P_J^{(8)}]g\rangle + \mathcal{O}(v^2)\;.
\end{equation}

\begin{figure}[h]
\includegraphics[width=0.48\textwidth]{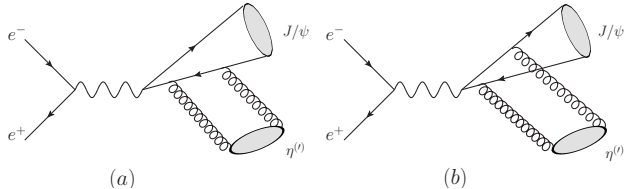}
\caption{The typical Feynman diagrams for electron-position to
$\eta^{(\prime)}$ $J/\psi$ transition. } \label{Fig:feynman}
\end{figure}

\begin{figure}[h]
\includegraphics[width=0.48\textwidth]{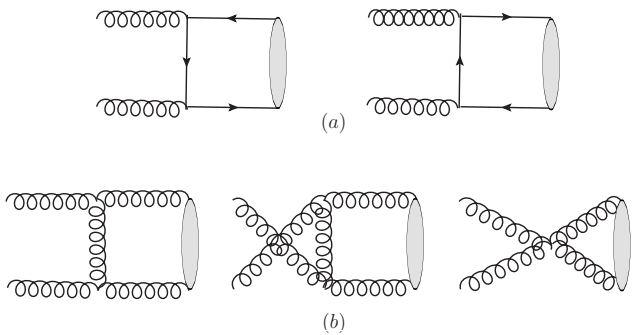}
\caption{The Feynman diagrams for two gluons  transiting to
different components of $\eta^{(\prime)}$: (a) to the meson's
quark component and (b) to its gluonium component.}
\label{Fig:feynman-2}
\end{figure}

Then let us consider the quark-level diagrams for the process $e^+e^-\to
\eta^{(\prime)} J/\psi$, which are depicted in
Figs.~\ref{Fig:feynman} and \ref{Fig:feynman-2}. With regard to the
conservation of parity and charge conjugation, the leading-order
contribution in $\alpha_s$ expansion comes from
Fig.~\ref{Fig:feynman} where the light meson $\eta^{(\prime)}$ is
emitted in the transition from two gluons. The color-singlet Fock state
dominates the hadronization of $J/\psi$ while the contribution from
color-octet Fock state
has a suppression factor of order $v^2\,\alpha_s$. In Fig.~\ref{Fig:feynman-2},
the relevant diagrams for two gluon transiting to different
components of $\eta^{(\prime)}$ are depicted.

The effective $\eta^{(\prime)} g^* g^*$ vertex from the meson's
quark component, as shown in Fig.~\ref{Fig:feynman-2}(a), can be
constructed as

\begin{equation}
{\cal M}^{(q)} \equiv -i \, F^{(q)}_{\eta^{(\prime)} g^* g^*}
(q_1^2, q_2^2, m_{\eta^{(\prime)}}^2) \, \delta_{a b} \,
\varepsilon^{\mu \nu \rho \sigma} \, \varepsilon^{a*}_\mu
\varepsilon^{b*}_\nu q_{1\rho} q_{2\sigma}, \label{eq:FFQ-1}
\end{equation}
where $q_1$ and $q_2$ are the momenta of two virtual gluons
respectively. $F^{(q)}_{\eta^{(\prime)} g^* g^*}$ represents the
transition form factor, and its expression is
\begin{eqnarray}
&&F^{(q)}_{\eta^{(\prime)} g^* g^*} (q_1^2, q_2^2,
m_{\eta^{(\prime)}}^2) =  \frac{2 \pi \alpha_s (\mu^2)}{ N_c}
\sum_{a=q,s}^{} f^{a}_{\eta^{(\prime)}}\int_0^1 dx \,\nonumber
\\&&~~\times \Phi^{(a)} (x,
\mu)  \left[ \frac{1} {x q_1^2 + \bar x q_2^2 - x \bar x
m_{\eta^{(\prime)}}^2 + i \epsilon} + (x \leftrightarrow \bar x)
\right],\label{eq:QFF-2}\nnb
\end{eqnarray}
here, the momentum exchange invariance  between $q_1$ and $q_2$
obviously holds. The factors $f^{a}_{\eta^{(\prime)}}$ are relevant
to the decay constants of $|\eta_q\rangle$ and $|\eta_s\rangle$
\cite{azizi}
\begin{eqnarray}
f^{q}_{\eta} &=& f_q\cos\phi\,,~~~~~~ f^{s}_{\eta} =
-f_s\sin\phi\,,\\
f^{q}_{\eta^{\prime}} &=& f_q\sin\phi\,,~~~~~~ f^{s}_{\eta^{\prime}}
= f_s\cos\phi\,.
\end{eqnarray}
We adopt the value of the decay constants of $\eta_q$ and
$\eta_s$ from Ref.~\cite{Feldman}
\[
 f_{q}=(1.07\pm0.02)f_\pi\,,
 ~~~~ f_{s}=(1.34\pm0.06)f_\pi\,,
 \]
with the  pion's decay constant $0.130\mathrm{GeV}$ \cite{BcPi}.

Furthermore,  the effective $\eta^{\prime} g^* g^*$ vertex can be
defined as
\begin{equation}
{\cal M}^{(g)} \equiv - i \, F^{(g)}_{\eta^{\prime} g^* g^*} \,
\delta_{a b} \, \varepsilon^{\mu \nu \rho \sigma} \,
\varepsilon^{a*}_\mu \varepsilon^{b*}_\nu q_{1\rho} q_{2\sigma} \,,
\label{eq:FFG-def}
\end{equation}
where the transition form factors $F^{(g)}_{\eta^{\prime} g^* g^*}$
have been calculated  in Ref.\cite{etagg}

\begin{eqnarray}
 &&F^{(g)}_{\eta^{\prime} g^* g^*} (q_1^2, q_2^2, m_{\eta^\prime}^2)
=\frac{4 \pi \alpha_s (\mu^2)}{Q^2} \, \frac{C}{2} \int_0^1 dx \,
\Phi^{(g)} (x, \mu)
\nonumber \\
&&  ~~~~~~\times\left [ \frac{x q_1^2 + \bar x q_2^2 - (1 + x \bar
x) m_{\eta^\prime}^2}
     {\bar x q_1^2 + x q_2^2 - x \bar x m_{\eta^\prime}^2 + i \epsilon}
- (x \leftrightarrow \bar x) \right ]\ , \label{eq:GFF-result1}
\end{eqnarray}
where $C=\sqrt 2 \, f_q \sin \phi + f_s \cos \phi$, and
$Q^2=|q_1^2+q_2^2|$. Note that the LC distribution amplitude of
the gluonium component is asymmetrical when we exchange the  momentum
fractions x and $\bar{x}$ in equation~(\ref{eq:gg}). Given the consideration,
 it can be seen that the form factor in equation~(\ref{eq:GFF-result1}) is
actually invariant  by exchange of the momenta of the incoming gluons.

Following the notation above, the amplitude  of $e^+e^-\to \eta
J/\psi$ at order  $\mathcal{O}(\alpha_s^4)$ can be obtained. The result
is presented in APPENDIX~\ref{app-amp}. To manipulate the trace and
matrix element square, the Mathematica software is employed with the
help of the packages FeynCalc\cite{feyncalc}, FeynArts\cite{arts},
and LoopTools\cite{Hahn}. The amplitudes are Ultra-Violet and
Infre-Red safe, and they can be calculated in four dimensions.

\section{production cross section\label{sec-num}}
In this section, we calculate the cross section  of  $e^+e^-\to
\eta^{(\prime)} J/\psi$. To compare with the experimental dada measured by
the BESIII and Belle collaborations, we vary the center-of-mass
energy  from $4$ {GeV} to $5.3$ {GeV}. In the numerical calculation,
the following values  of input parameters are adopted \cite{pdg}
\[
 m_{J/\psi}=3.096\,\mathrm{GeV},~m_{\eta}=547.8\,\mathrm{MeV},
 ~m_{\eta^\prime}=957.7\,\mathrm{MeV}\,,
 \]
\[
 m_c=1.40\,\mathrm{GeV}\,,~~~~\alpha=1/127,
 ~~~~\Gamma_{ee}^{J/\psi}=5.55~\mathrm{keV}\ .
 \]

The radial wave function squared at the origin of the $J/\Psi$ is
extracted from its leptonic width at leading-order of
$\alpha_s$\;\cite{Qiao:2011}, i.e.,
\begin{eqnarray} |R(0)|_{J/\Psi}^2=\frac{m^2_{J/\Psi }
\Gamma (J/\Psi\rightarrow e^+ e^-)}{4\alpha^2e_{c}^2}\ .
\end{eqnarray}
Then the $J/\Psi$'s Schr\"{o}dinger wave function squared at the
origin can be  obtained from
$|\psi(0)|_{J/\Psi}^2=|R(0)|_{J/\Psi}^2/4\pi$, with numerical value
of $0.0446~\mathrm{GeV}^{3}$ .

Thus the cross section for the process $e^+e^- \to \eta J/\psi$ can
be obtained. At center-of-mass energy $\sqrt{s}=4.009\mathrm{GeV}$,
the numerical result is $34.6\mathrm{pb}$. The result is consistent
with the measurement $32.1\pm2.8\pm1.3\mathrm{pb}$ obtained by the BESIII
Collaboration. We can conclude that the two-gluons transition
dominates the creation of $\eta$ and higher order contribution should be
suppressed by a small K factor.
\begin{figure}
\includegraphics[width=0.45\textwidth]{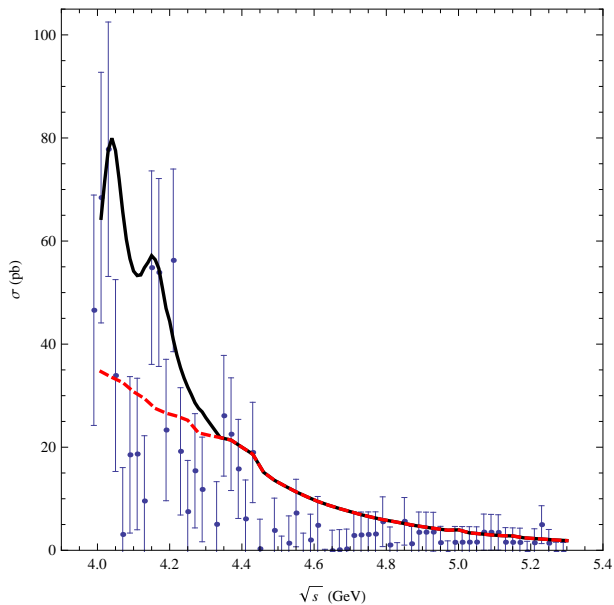}
\hspace{5mm} \caption{The cross section of $e^+e^-\to\eta J/\psi$
versus center-of-mass energy $\sqrt{s}$. The dashed line is the
cross section including the contributions shown in Fig.~\ref{Fig:feynman}, while the
solid line is the result after considering the resonance state
effects, for which we adopt the effective coupling constants
$g_{\psi(X)c\bar{c}}~g_{\psi(X)e^+e^-}=\sqrt{5}e^2/100$. And the
data is measured by the Belle collaboration~\cite{Belle2}.} \label{Fig:eta-cs}
\end{figure}

For the comparison with the data measured by the Belle
Collaboration at $\sqrt{s}$ from  $4$ {GeV} to $5.3$ {GeV}, the
production mechanism presented in Figs.~\ref{Fig:feynman} and
\ref{Fig:feynman-2} is incomplete in specific energy regions. In fact, we should
take into account the resonance-state effects or the
open-charm effects when $\sqrt{s}$  is near the
resonance states which are strongly coupling to $\eta J/\psi$. In
the regions when $\sqrt{s}$ is near the $\psi(4040)$ and
$\psi(4160)$ peaks, the virtual D-meson loop effect becomes
important\cite{Wang:2011yh,mlc,IML,IML-2}. Here, we adopt the effective coupling
constants $g_{\psi(X)c\bar{c}}$ and $g_{\psi(X)e^+e^-}$  with
$X=4040, 4160$ to phenomenologically describe the resonance effects.
We can write the resonance $\psi(X)$ transition amplitude as
\begin{equation}
    M=\bar{v}_{e^-}\gamma^\alpha
    u_{e^+}\frac{g_{\psi(X)c\bar{c}}g_{\psi(X)e^+e^-}
    (-g^{\alpha\beta}+\frac{p^\alpha
    p^\beta}{m^2_{\psi(X)}})}{e(s-m^2_{\psi(X)}+im_{\psi(X)}\Gamma_{\psi(X)})}M_\beta\,,
\end{equation}
where p denotes the momentum of the virtual photon in
Fig.~\ref{Fig:feynman},  $e$ is related to
 the electromagnetic coupling constant with $e^2=4\pi \alpha$, $m_{\psi(X)}$ and
$\Gamma_{\psi(X)}$ represent the mass and the decay width of $\psi(X)$
respectively, while $M_\beta$ can be obtained through a virtual
photon emitted to $\eta J/\psi$ at order
$\mathcal{O}(\alpha_s^4)$.

We plot our results and experimental data together in
Fig.~\ref{Fig:eta-cs}. There the $\sqrt{s}$ dependence of the cross section of $e^+e^-\to\eta
J/\psi$ is presented.
The dashed line is the contribution from Fig.~\ref{Fig:feynman},
while the solid line is the contribution after considering the
resonance state effects. We find that the cross section at the
$\psi(4040)$ and $\psi(4160)$ peaks can be well explained when we
fix the effective coupling constants
$g_{\psi(X)c\bar{c}}~g_{\psi(X)e^+e^-}=\sqrt{5}e^2/100$.

In comparison with the result based on open charm effect in
Ref.~\cite{Wang:2011yh}, we find that our results are consistent with
those when the phase parameters are chosen as $(\theta,\beta,\phi)=(0,0,0)$.

Finally, to allow the experimental search for a signal of gluonium component in
$\eta^\prime$ we have calculated the cross sections with and without such contribution.
The results are  given in
Table~\ref{Tab:cs-prime}. There we assume, as in  Ref.~\cite{mix-ang}, that  $26\%$ of
gluonic admixture  exists in $\eta^\prime$,i.e. $\mathrm{sin}\phi_G^2=0.26$. The numerical results
show that such contribution from the gluonium component can affect the total cross
section by up to a $17.0\%$ variation. This possibility can be checked in the upcoming Belle experiment.
We also find that the cross section decreases rapidly with increasing scattering energy.

\begin{widetext}
\begin{center}
\begin{table}[h]
\caption{\label{Tab:cs-prime} The cross sections (in $\mathrm{pb}$)  of $e^+e^- \to
\eta^\prime J/\psi$ at different center-of-mass energies $\sqrt{s}$.
Here  $\sigma_{w}$ and $\sigma_{w/o}$ represent the cross sections
with and without the gluonium component contribution respectively. }
\vskip 0.6cm
\begin{tabular}{|c|c|c|c|c|c|c|c|c|c|c|c|}
  \hline
~~$\sqrt{s}(\mathrm{GeV})$ &4.3 & 4.4&4.5 &4.6 &4.7 &4.8 &4.9 &5.0
&5.1 &5.2 &5.3
\\\hline
 ~~$\sigma_{w/o}$ &41.2&
29.2& 19.7& 14.5&11.1&8.84& 6.78& 5.37& 4.29& 3.46&2.79\\\hline
 ~~$\sigma_{w}$& 34.1& 24.2& 16.4& 12.6& 9.9&8.23& 6.51& 5.22& 4.24&
 3.45& 2.82\\
  \hline
   $\frac{|\sigma_{w}-\sigma_{w/o}|}{\sigma_{w/o}}$ & 17.0\% &16.8\%
    &16.3\% &13.4\% &10.0\% &6.9\% &3.9\% &2.7\% &1.3\% &0.2\% &1.1\% \\
  \hline
\end{tabular}
\end{table}
\end{center}
\end{widetext}

\section{Conclusions\label{sec-con}}

Motivated by the large cross section of $e^+e^-\to \eta J/\psi$
process observed by the BESIII and Belle Collaborations, we have evaluated
this process in the framework of NRQCD. We find that the cross
section at $\sqrt{s}=4.009$ {GeV} is $34.6\mathrm{pb}$, which is in
agreement with the BESIII data  when its uncertainty is taken into
account. And our results  can also explain the energy dependence
measured by the Belle Collaboration. Inspired by this calculation, we
have also estimated the cross section for $e^+e^- \to \eta^\prime J/\psi$
process, where the contribution of two gluons transiting to $\eta^\prime$ is treated
within the LC approach.

The electronic production of $\eta^{(\prime)}J/\psi$ has a crucial
phenomenological significance.  Color-singlet
 transitions dominate the formation of the accompanying $J/\psi$
 and color-octet Fock states are suppressed by a factor of $v^2\alpha_s$.
 This clear hadronization
structure of $J/\psi$ allows us to focus on the interesting mixing effect
between $\eta$ and $\eta^\prime$
and differentiate the quark and gluonoium components
of the $\eta^\prime$.  Concerning the $\eta$, the comparison between the experimental data
for
 $e^+e^-\to \eta J/\psi$
and our calculations confirm the traditional view that
the probability of a gluonoium component in $\eta$
is negligible.

The cross section dependence on the center-of-mass energy $\sqrt{s}$
for $e^+e^- \to \eta J/\psi$ will provide a platform to investigate
the properties of resonance states such as $\psi(4040)$ and
$\psi(4160)$. By measuring the cross section  and comparing
it with the contribution from Fig.~\ref{Fig:feynman}, one can also
hunt the signals for the potential resonances strongly coupling to
$\eta J/\psi$ and(or) measure their branching ratios.

Finallly, we have taken the gluonoium component into account for $e^+e^- \to
\eta^\prime J/\psi$. The gluonium component has been investigated in many
 processes where a heavy meson decays to $\eta^\prime$ and a light meson.
 Here we have studied it in
electroproduction. In the energy scan region of the Belle experiment,  we
calculated the contributions from both  quark and
gluonium components. Numerical results show that the gluonium
component contributions would decrease the total cross section by up to
$17\%$, an effect that can be detected in the upcoming experiment.

\hspace{2cm}

{\bf Acknowledgements}:

R. Z. thanks Prof. Feng Yuan for useful discussions. This work was
supported in part by the National Natural Science Foundation of
China(NSFC) under the grants 10935012, 11121092, 11175249, and
 11375200.

\hspace{2cm}

\appendix

\section{Amplitude of $e^+e^-\to \eta J/\psi$\label{app-amp}}
In this appendix, the amplitude of $e^+e^-\to \eta J/\psi$  is given
at $\mathcal{O}(\alpha_s^4)$ accuracy in the framework of
nonrelativistic QCD. In the formula below, $s$ is the center-of-mass energy
squared,  $x$ and $\bar{x}$ are the momentum-fractions of the two
partons inside the $\eta$ meson component, $\Phi(x)$ is the
light-cone distribution amplitude of $\eta_q$, $\varepsilon$ is the
polarization vector of $J/\psi$, and $ \text{B}_0 $, $ \text{C}_0 $,
and $ \text{D}_0 $ are scalar Passarino-Veltman integrals defined in
Ref.~\cite{Hahn}. Besides, we indicate the amplitudes of the processes (a) and (b)
in Fig.~\ref{Fig:feynman} with
 $\mathcal{M}_a$ and $\mathcal{M}_b$, respectively. The total
 amplitude can be written as $\mathcal{M}=2\mathcal{M}_a+2\mathcal{M}_b$.
\begin{widetext}
\begin{eqnarray}
   && \mathcal{M}_a=\int_0^1 dx \,\frac{32 \sqrt{6} \pi  \sqrt{m_c}
   \epsilon_{\lambda\mu\nu\rho}\varepsilon^\mu
   p_c^\nu p_q^\rho \bar{v}_e\gamma^\lambda u_e \Phi(x)
   \psi(0)_{J/\Psi}C_F\alpha \alpha _s^2}{9 s
    (-8 m_c^2  (m_{\eta }^2+s )+16 m_c^4+ (s-m_{\eta }^2 ){}^2
    )}\sum_{a=u,d,s}^{} f^{a}_{\eta}\times\{\frac{m_{\eta }^4
    (-4 m_c^2+m_{\eta }^2-s )
   }{-4
   m_c^2+m_{\eta }^2+s}D_1
    \nonumber\\
   &&  -\frac{ (m_c^2
    ((16-x (4 x+11)) m_{\eta }^2+s (16-13
   x) )+4 (7 x-8) m_c^4+(x-1)  (3 s x m_{\eta }^2+(x+2)
   m_{\eta }^4+2 s^2 ) )}{(x-1)  (-4 m_c^2+m_{\eta
   }^2+s )}C_1
   \nonumber\\
   && +\frac{ (4 m_c^2-m_{\eta }^2+s )  (x (x+1) m_{\eta }^2-m_c^2 )
  }{x  (-4 m_c^2+m_{\eta }^2+s )} C_2+2 m_{\eta }^2 C_3
  +\frac{ (-4 m_c^2+m_{\eta }^2+3 s ) B_0 (\frac{1}{2}  (-2 m_c^2+m_{\eta
   }^2+s ),0,m_c^2 )}{(x-1)  (-4 m_c^2+m_{\eta }^2+s )} \nonumber\\
   &&-\frac{x  (-4 m_c^2+m_{\eta }^2+3
   s ) B_0 ((1-2 x) m_c^2+\frac{1}{2} x  ((2 x-1) m_{\eta }^2+s ),0,0 )}
   {(x-1)  (-4
   m_c^2+m_{\eta }^2+s )} +2 B_0 ((x-1)^2 m_{\eta }^2,0,0 )\nonumber\\
   &&+\frac{ (-4 (x-1) m_c^2+(x-1) m_{\eta }^2+3 s x+s ) B_0 ((1-2 x)
   m_c^2+\frac{1}{2} x  ((2 x-1) m_{\eta }^2+s ),0,m_c^2 )}{x  (-4 m_c^2
   +m_{\eta
   }^2+s )} \nonumber\\
   &&+\frac{B_0 (m_c^2,0,0 )  (-4 m_c^2+m_{\eta }^2-s )}{x  (-4 m_c^2
   +m_{\eta
   }^2+s )}-2 B_0 (x^2 m_{\eta }^2,0,0 ) +(x \leftrightarrow \bar x)\}\,,
\end{eqnarray}
\begin{eqnarray}
   && \mathcal{M}_b=-\int_0^1 dx \,\frac{32 \sqrt{6} \pi  \sqrt{m_c}
   \epsilon_{\lambda\mu\nu\rho}\varepsilon^\mu
   p_c^\nu p_q^\rho \bar{v}_e\gamma^\lambda u_e \Phi(x)
   \psi(0)_{J/\Psi}C_F\alpha \alpha _s^2}{9 s x^2  (16 m_c^4-8  (m_{\eta
   }^2+s ) m_c^2+ (s-m_{\eta }^2 ){}^2 )}\sum_{a=u,d,s}^{} f^{a}_{\eta}
   \times\{
   \frac{4 x  (4 m_c^2+2 m_{\eta }^2-s ) m_c^2}{x-1}D_2\nonumber\\
   &&+\frac{4 x
   m_{\eta }^2  (4 m_c^2+ (2 x^2+2 x+1 ) m_{\eta }^2-s ) m_c^2}{4 m_c^2
   +m_{\eta }^2-s}D_3+4 x  m_{\eta
   }^2C_3-\frac{4
    (4 (3 x-2)
   m_c^2+(3-2 x) x m_{\eta }^2+s (2-3 x) ) m_c^2}{(x-1)  (4 m_c^2
   +m_{\eta }^2-s )}C_4\nonumber\\
   &&-\frac{4 (x-1)
     (-4 m_c^2+(2 x+1) m_{\eta }^2+s ) m_c^2}{4 m_c^2+m_{\eta
   }^2-s}C_5+\frac{2
   x  (16 m_c^4-8  (s-2 x m_{\eta }^2 )
   m_c^2+ (4 x^3-1 ) m_{\eta }^4+s^2-4 s x m_{\eta }^2 ) m_c^2}
   {(x-1)  (-4 m_c^2-m_{\eta
   }^2+s )}D_4\nonumber\\
   &&+\frac{4   (-4 (x-1) m_c^4+ ( (2 x^2+3 x-1 ) m_{\eta }^2+s (x-1) )
   m_c^2+x m_{\eta }^2  (m_{\eta }^2-s ) )}{-4 m_c^2-m_{\eta }^2
   +s}C_1-\frac{2 x
    (4 m_c^2+(2 x-1) m_{\eta }^2-s )}{x-1}C_7\nonumber\\
 &&-\frac{2
     (-8  (2 x^2-5 x+2 )
   m_c^4+2  (x  (4 x^2-1 ) m_{\eta }^2+s  (2 x^2-7 x+2 ) )
   m_c^2+x  (s-m_{\eta
   }^2 )  ((1-2 x) m_{\eta }^2+s ) )}{(x-1)  (-4 m_c^2
   -m_{\eta }^2+s )}C_8\nonumber\\
 &&+\frac{2 x
    (-4 m_c^2-m_{\eta }^2+s )}{x-1}C_6+\frac{2
     (16
   (3 x-1) m_c^4+4  ( (-2 x^2+4 x+1 ) m_{\eta }^2+s
   -4 s x ) m_c^2+x  (s-m_{\eta
   }^2 ){}^2 )}{(x-1)  (4 m_c^2+m_{\eta }^2-s )}C_9\nonumber\\
 &&-\frac{x
   (-32 m_c^6+16  (m_{\eta }^2+2 s )
   m_c^4-2  (3 m_{\eta }^4+4 s (x-1) m_{\eta }^2+5 s^2 ) m_c^2+
   (s-m_{\eta }^2 ){}^2  ((2
   x-1) m_{\eta }^2+s ) )}{-4 m_c^2-m_{\eta }^2+s}D_5\nonumber\\
 &&+\frac{2 x m_{\eta }^2  ((8-16
   x) m_c^4+2  ( (2 x^2-2 x+1 ) m_{\eta }^2+s (4 x-1) ) m_c^2-x  (s-m_{\eta
   }^2 ){}^2 )}{4 m_c^2+m_{\eta }^2-s}D_6 +(x \leftrightarrow \bar x)\}\,,
\end{eqnarray}

 where
 \begin{eqnarray}
&&D_1=\text{D}_0 (m_c^2,m_{\eta }^2,f_1,f_3,x^2 m_{\eta
}^2,0,0,m_c^2,0 )
 \,,\nonumber\\
 &&D_2=\text{D}_0 (m_c^2,f_3,m_c^2,f_3,s,m_{\eta
   }^2,0,0,m_c^2,m_c^2 ) \,,\nonumber\\
 &&
   D_3=\text{D}_0 (m_c^2,x^2 m_{\eta }^2,\bar{x}^2 m_{\eta }^2,f_3,f_1,
   m_{\eta }^2,0,0,0,m_c^2 )\,,\nonumber\\
   &&D_4=\text{D}_0 (m_c^2,x^2 m_{\eta }^2,f_2,s,f_1,f_3,0,0,m_c^2,
 m_c^2 )\,,\nonumber\\
 &&D_5= \text{D}_0 (f_1,f_2,m_c^2,f_3,s,\bar{x}^2 m_{\eta }^2,0,0,
 m_c^2,m_c^2 ) \,,\nonumber\\
 &&D_6= \text{D}_0 (x^2 m_{\eta
   }^2,f_3,m_c^2,\bar{x}^2 m_{\eta }^2,f_2,m_{\eta }^2,0,0,0,m_c^2 )
   \,,\nonumber
   \end{eqnarray}
\begin{eqnarray}
 &&C_{1}=\text{C}_0 (f_3,\bar{x}^2 m_{\eta }^2,f_1,0,m_c^2,0 )\,,
 \nonumber\\
 && C_2=\text{C}_0 (m_c^2,x^2 m_{\eta }^2,f_1,0,0,m_c^2 )\,,
 C_9=C_2|_{x\rightarrow1}\,,\nonumber\\
 &&
   C_3=\text{C}_0 (m_{\eta
   }^2,x^2 m_{\eta }^2,\bar{x}^2 m_{\eta }^2,0,0,0 )\,,\nonumber\\
 &&
   C_4=\text{C}_0 (m_c^2,s,f_3,m_c^2,0,m_c^2
   )\,,\nonumber\\
 &&
   C_5=\text{C}_0 (m_c^2,\bar{x}^2 m_{\eta }^2,f_2,0,0,m_c^2 )\,,
   \nonumber\\
 &&C_6=\text{C}_0 (m_c^2,s,f_3,0,m_c^2,m_c^2
   )\,,\nonumber\\
 &&C_7=\text{C}_0 (f_3,x^2 m_{\eta }^2,f_2,m_c^2,0,0 )\,,
 \nonumber\\
 &&
 C_8=\text{C}_0 (s,f_2,f_1,0,m_c^2,m_c^2 )\,,\nonumber\\
 &&
   f_1=(\bar{x}-x) m_c^2+\frac{1}{2} x  ((x-\bar{x}) m_{\eta }^2+s )\,,
   \nonumber\\
 &&
   f_2=\frac{1}{2}  (2(x-\bar{x})
   m_c^2-\bar{x}  ((\bar{x}-x) m_{\eta }^2+s ) )\,,\nonumber\\
 &&f_3=\frac{1}{2}  (-2 m_c^2+m_{\eta
   }^2+s )\,.
\end{eqnarray}
\end{widetext}

 \hspace{2cm}


\begin{thebibliography}{99}

\bibitem{HQW}
      N.~Brambilla et al. (Quarkonium Working Group), Eur.\ Phys.\ J.\ C {\bf 71}, 1534 (2011).
\bibitem{kzhu}
     K.~Zhu,  32th International Symposium on Physics in Collision (2012),
Strbske Pleso, Slovakia, arXiv:1212.2169.
\bibitem{Belle}
   B.~Aubert et al. (BABAR Collaboration), Phys.\ Rev.\ Lett.\ {\bf 95},
142001 (2005); C.~Z.~Yuan et al. (Belle Collaboration), Phys.\ Rev.\
Lett.\ {\bf 99}, 182004 (2007); X.~-L.~Wang et al. (Belle
Collaboration), Phys.\ Rev.\ Lett.\ {\bf99}, 142002 (2007).
\bibitem{Coan:2006rv}
  T.~E.~Coan {\it et al.}  (CLEO Collaboration),
  Phys.\ Rev.\ Lett.\  {\bf 96}, 162003 (2006).
\bibitem{BesIII}
     M.~Ablikim et al. (BESIII Collaboration), Phys.\ Rev.\ D {\bf 86},
     071101 (2012).
\bibitem{Belle2}
     X.~-L.~Wang et al. (Belle Collaboration),
     Phys.\ Rev.\ D {\bf87},
     051101 (2013).
\bibitem{Wang:2011yh}
  Q.~Wang, X.~-H.~Liu, and Q.~Zhao,
  Phys.\ Rev.\ D {\bf 84}, 014007 (2011).

\bibitem{exp}
   B.~Aubert et al. (BABAR Collaboration), Phys.\ Rev.\ D
{\bf 72}, 031101 (2005) ; K.~Abe et al. (Belle Collaboration), Phys.\
Rev.\ D {\bf70}, 071102 (2004).
\bibitem{theore}
   K.~-Y.~Liu, Z.~-G.~He, and K.~-T.~Chao, Phys.\ Lett.\ B
{\bf557}, 45(2003); E.~Braaten and J.~Lee, Phys.\ Rev.\ D {\bf67},
054007 (2003); K.~Hagiwara,
E.~Kou, and C.~-F.~Qiao, Phys.\ Lett.\ B {\bf570},  39 (2003).
\bibitem{NLO}
  Y.~-J.~Zhang, Y.~-J.~Gao, and K.~-T.~Chao,
  Phys.\ Rev.\ Lett.\  {\bf 96}, 092001 (2006); B.~Gong and J.~-X.~Wang, Phys.\ Rev.\ D
{\bf 77}, 054028 (2008).
\bibitem{He:2007te}
  Z.~-G.~He, Y.~Fan, and K.~-T.~Chao,
  Phys.\ Rev.\ D {\bf 75}, 074011 (2007).
\bibitem{Bramon:1997va}
  A.~Bramon, R.~Escribano, and M.~D.~Scadron,
  Eur.\ Phys.\ J.\ C {\bf 7}, 271 (1999).
\bibitem{decay1}
G.~Ricciardi, Phys.\ Rev.\ D {\bf 86}, 117505 (2012);
  C.~D.~Donato, G.~Ricciardi, and I.~Bigi,
  Phys.\ Rev.\ D {\bf 85}, 013016 (2012).
 \bibitem{decay2}
Y.~-Y.~Fan, W.~-F.~Wang, S.~Cheng, and Z.~-J.~Xiao,
  Phys.\ Rev.\ D {\bf 87}, 094003 (2013); X.~Liu, H.~-N.~Li, and Z.~-J.~Xiao, Phys.\ Rev.\ D {\bf 86}, 011501 (2012).
 \bibitem{decay3}
  S.~Dubnicka, A.~Z.~Dubnickova, M.~A.~Ivanov, and A.~Liptaj,
  Phys.\ Rev.\ D {\bf 87}, 074201 (2013).
 \bibitem{prod1}
  A.~I.~Ahmadov, D.~G.~Kostunin, and M.~K.~Volkov,
  Phys.\ Rev.\ C {\bf 87}, 045203 (2013).
  \bibitem{Feldman}
   T.~Feldmann, P.~Kroll, and B.~Stech, Phys.\ Rev.\ D {\bf 58}, 114006 (1998);
   T.~Feldmann, Int.\ J.\ Mod.\ Phys.\ A {\bf 15}, 159 (2000).
\bibitem{KLOE}
 F.~Ambrosino et al. (KLOE Collaboration), Phys.\ Lett.\ B {\bf 648}, 267 (2007).
\bibitem{cgge}
 F.~D.~Fazio and M.~R.~Pennington, J.\ High Energy Phys.\ 07 (2000) 051.
\bibitem{Ball}
P.~Ball, J.\ High Energy Phys.\ 01 (1999) 010.
\bibitem{etagg}
 A.~Ali and A.~Y.~Parkhomenko, Phys.\ Rev.\ D {\bf 65}, 074020 (2002); T.~Muta and M.~-Z.~Yang,
  Phys.\ Rev.\ D {\bf 61}, 054007 (2000).
\bibitem{Bratten}
   G.~T.~Bodwin, E.~Braaten, and G.~P.~Lepage, Phys.\ Rev.\
D {\bf 51}, 1125 (1995).
\bibitem{Qiao:2009}
   C.~-F.~Qiao, L.~-P.~Sun, and P.~Sun,  J.\ Phys.\ G {\bf 37}, 075019 (2010);
  C.~-F.~Qiao, L.~-P.~Sun, and R.~-L.~Zhu,
   J.\ High Energy Phys.\ 08 (2011) 131.
\bibitem{azizi}
K.~Azizi, R.~Khosravi, and F.~Falahati, Phys.\ Rev.\ D {\bf 82}, 116001
(2010).
\bibitem{BcPi}
   C.~-F.~Qiao, P.~Sun, D.~Yang, and R.~-L.~Zhu,
  Phys.\ Rev.\ D {\bf 89}, 034008 (2014).
\bibitem{feyncalc}
R.~Mertig, M.~Bohm, and A.~Denner, Comput.\ Phys.\ Commun, {\bf 4}, 345
(1991).
\bibitem{arts}
   T.~Hahn, Comput.\ Phys.\ Commun, {\bf 140}, 418 (2001).
\bibitem{Hahn}
T.~Hahn and M.~Perez-Victoria, Comput.\ Phys.\ Commun, {\bf 118}, 153
(1999).
\bibitem{pdg}
J.~Beringer et al. (Particle Data Group), Phys.\ Rev.\
D {\bf 86}, 010001 (2012).
\bibitem{Qiao:2011}
 C.~-F.~Qiao and R.~-L.~Zhu, Phys.\ Rev.\ D {\bf 87}, 014009 (2013).
\bibitem{mlc}
  Z.~-G.~Guo, S.~Narison, J.~-M.~Richard, and Q.~Zhao,
  Phys.\ Rev.\ D {\bf 85}, 114007 (2012);  G.~Li, X.~-H.~Liu, Q.~Wang, and Q.~Zhao,
  Phys.\ Rev.\ D {\bf 88}, 014010 (2013); Q.~Wang, C.~Hanhart, and Q.~Zhao,
  Phys.\ Rev.\ Lett.\  {\bf 111}, 132003 (2013).
\bibitem{IML}
  D.~-Y.~Chen, X.~Liu, and T.~Matsuki,
  Phys.\ Rev.\ D {\bf 87}, 054006 (2013).
\bibitem{IML-2}
      F.~-K.~Guo, C.~Hanhart, and U.~-G.~Meissner,
  Phys.\ Rev.\ Lett.\  {\bf 103}, 082003 (2009);
 {\bf 104}, 109901E (2010).
\bibitem{mix-ang}
E.~Kou and A.~I.~Sanda, Phys.\ Lett.\ B {\bf 525}, 240 (2002).

\end{thebibliography}
\end{document}